\documentclass[iop]{emulateapj}
\usepackage{graphicx}
\usepackage{graphics}
\usepackage{amsmath}
\usepackage{multirow}
\usepackage{color}
\usepackage{xcolor}

\usepackage{float}

\def\kmskpc{{\rm\,km\, \,s^{-1} \, {kpc}^{-1}}}

\def\mathnew{\mathsurround=0pt}   
\def\simov#1#2{\lower .5pt\vbox{\baselineskip0pt  
    \lineskip-.5pt\ialign{$\mathnew#1\hfil##\hfil$\crcr#2\crcr\sim\crcr}}}

\def\'#1{\ifx#1i{\accent"13\i}\else{\accent"13#1}\fi}

\begin{document}

\shorttitle{Survival of High-Altitude Open Clusters within the Milky Way Tides} 
\shortauthors{Martinez-Medina et al. 2016}

\title{On the Survival of High-Altitude Open Clusters within the Milky Way Galaxy Tides}
\author{L.A. Martinez-Medina, B. Pichardo,  A. Peimbert \& E. Moreno}
\affil{Instituto de Astronom\'ia, Universidad Nacional Aut\'onoma de M\'exico, A.P. 70--264, 04510, M\'exico, D.F., M\'exico; \texttt{lamartinez@astro.unam.mx}}


\begin{abstract}
It is a common assumption that high-altitude open clusters live longer
compared with clusters moving close to the Galactic plane. This is
because at high altitudes, open clusters are far from the disruptive
effects of in-plane substructures, such as spiral arms, molecular
clouds and the bar. However, an important aspect to consider in this
scenario is that orbits of high-altitude open clusters will eventually
cross the Galactic plane, where the vertical tidal field of the disk
is strong. In this work we simulate the interaction of open clusters
with the tidal field of a detailed Milky Way Galactic model at
different average altitudes and galactocentric radii. We find that the
life expectancy of clusters decreases as the maximum orbital altitude
increases and reaches a minimum at altitudes of approximately 600
pc. Clusters near the Galactic plane live longer because they do not
experience strong vertical tidal shocks from the Galactic disk; then,
for orbital altitudes higher than 600 pc, clusters start again to live
longer due to the decrease in the number of encounters with the
disk. With our study, we find that the compressive nature of the tides
in the arms region and the bar have an important role on the survival
of small clusters by protecting them from disruption: clusters inside
the arms can live up to twice as long as those outside the arms at
similar galactocentric distance.

\end{abstract}


\section{Introduction} \label{sec:intro}
Stars in the disk of our Galaxy seem to have been primarily formed in
open clusters. Newly born open clusters originate on the thin
molecular gas layer of the Galaxy \citep{2008ApJ...685L.125D}; most
originate within a galactocentric distance of 10.5 kpc and closer than
180 pc from the disk plane
\citep{2002A&A...389..871D,2012AstL...38..519G}. Less than 10\%
survive their embedded stage as gravitationally bound systems
\citep{2003ARA&A..41...57L}; the ones that survive as clusters, with
ages ranging between a few million years and approximately 10 Gyr, are
located between 5 and 20 kpc from the Galactic center.
\citep{1970ApJ...160..811F,1993ApJ...406..501N,1995ApJ...447L..95K,2002ApJ...581..258F,2002ASPC..273....7V,2006ApJ...643.1011P,2012MNRAS.419.1860D,2014ApJ...793..110M,2014A&A...564L...9R};
Albeit the majority are radially confined, and at very low altitudes
away from the Galactic disk plane, numerous clusters \citep[$\sim
  13\%$;][]{2008ApJ...685L.125D} are found at great altitudes; some of
those far beyond 200 pc and up to a few kpc.

Interesting scenarios on the origin of high-altitude clusters have
been proposed such as: capture from satellite galaxies, formation in
situ on high-altitude molecular clouds
\citep{1977MNRAS.180..709W,2008ApJ...685L.125D,2009MNRAS.399.2146W,2010MNRAS.407.2109V,2013MNRAS.434..194D},
and smoother dynamical intrinsic processes
\citep{Quillen2002,Quillen2014}. Based on the last idea, in a previous
paper \citep{2016ApJ...817L...3M}, we explored in detail the spiral
arms as the mechanism to induce the clusters lifting; this mechanism
was disregarded in the past based on studies with simple models of
very weak spiral arms (i.e., low amplitude and/or pitch angle). Those
models were characterized by having highly idealized potentials, only
meant to facilitate analytical calculations to understand better the
nature of the spiral arms \citep{1958ApJ...127...17S,Wielen1977}.

In their path through the Galaxy, stellar clusters undergo disruptive
stages that dissolve many of them. Lot of work has been devoted to
theoretical and observational studies on the evolution and survival of
clusters in general ---most of them globular clusters---
\citep[e.g.][]{1972ApJ...176L..51O,1997ApJ...474..223G,2003MNRAS.340..227B,2007MNRAS.376..809G,2008MNRAS.389L..61H,2010MNRAS.407.2241K,2010MNRAS.409..305L,2011MNRAS.413.2509G,2011MNRAS.414.1339K,2011MNRAS.418..759R,2012MNRAS.419.3244B,2012MNRAS.424.2614S,2013MNRAS.436.3695R,2013MNRAS.429.1066S,2014MNRAS.441..150B,2014MNRAS.440L.116S,2015MNRAS.448.3416R,2016arXiv160802309R}. The
first disruptive stage, related to two-body relaxation processes, is
known as the expansion phase
\citep{2000ApJ...535..759T,2003MNRAS.340..227B,2011MNRAS.413.2509G,2012ApJ...756..167M}. The
second stage is known as the evaporation phase
\citep{2011MNRAS.413.2509G}; in this period the tidal effects of the
host galaxy become of great importance for the endurance of clusters
as entities; several factors intervene in their evaporation such as
the characteristics of the orbit
\citep{2003MNRAS.340..227B,2014MNRAS.442.1569W}, depending for example
on their distance to the Galactic center \citep{2012ApJ...756..167M}
and the mass and size of the Galactic disk
\citep{2014ApJ...784...95M}.

In particular, in the case of disk clusters (called open clusters in
the Milky Way Galaxy), several contributions have to be taken into
account to explain the observed short dissolution times.  Structures
that have been considered to influence the survival expectations
include bars, spiral arms (either steady or transient), and molecular
clouds
\citep{2006MNRAS.371..793G,2006A&A...455L..17L,2007MNRAS.376..809G,2012MNRAS.419.3244B}. For
a recent detailed study on the case of the Sun's birth cluster, see
\citet{2016MNRAS.457.1062M}.

The strongest tides encountered by disk clusters occur when they go
through high density regions of the disk like giant molecular clouds
\citep{2006MNRAS.371..793G} or spiral arms
\citep{2007MNRAS.376..809G,2011MNRAS.414.1339K}. One of the strongest
tides for massive stellar clusters is the one exerted by the disk for
clusters that experiment large vertical excursions \citep[see][for a
  study on the effect on globular clusters]{1997ApJ...474..223G}.
Although, traditionally, open (or disk) clusters are assumed to have
experienced only small excursions away from the Galactic plane, this
is not the case for over 10\% of them.

On the other hand, since the 1950's, Oort drew attention to the fact
that there is a lack of old open clusters (older than 1 Gyr) in the
Galaxy; this cannot be explained by stellar evolution alone, which
would make it difficult to observe when they are older. This
discrepancy is not only seen in the Milky Way Galaxy, it can also be
seen, for example, in the galaxy M51
\citep{2005A&A...441..949G,2005A&A...429..173L}. The discrepancy
between the number of observed and predicted old open clusters implies
that a significant fraction must have been destroyed by different
processes
\citep{1971A&A....13..309W,2005A&A...441..117L,2007MNRAS.376..809G}.

The majority of open clusters in the Galaxy, travel as a unit all
their (generally short) lifetimes within the thinner disk in cold
(approximately gas-like) orbits. For those clusters it is a fact that
they cannot be affected by tides produced by the disk because they do
not suffer gravitational encounters with the disk. On the other hand,
in the case of the oldest open clusters ---that tend to be the ones
that present the largest excursions away from the plane---, the disk
itself is source of important tides able to destroy them as we will
show in this work.

Finally, important advances have been done in this field thanks to
observational and theoretical dynamical studies (most of them applied
to globular clusters). From the theoretical point of view, two
important complementary approaches have been employed to simulate a
disk galaxy: steady potentials
\citep[e.g.,][]{2014MNRAS.437..916G,2014MNRAS.444...80O,2016MNRAS.455..596C}
and $N$-body simulations
\citep[e.g.,][]{2006MNRAS.371..793G,2013MNRAS.436.3695R,2016arXiv160802309R}. While
$N$-body simulations provide the only way to study galaxy and clusters
evolution self-consistently, steady potentials are fully adjustable,
fast, able to represent a given galaxy (to the best of our
observational knowledge of it), and provide orbital details that
$N$-body simulations cannot do. Due to the nature of the problem we
are tackling we will employ a very detailed steady potential to better
represent the Milky Way Galaxy. In this work we focus on the
disruption of open clusters; both the formation rate and the initial
mass function of those clusters are beyond the scope of this paper.

This paper is organized as follows. In Section \ref{tides} we describe
the procedure that has been followed to build the Galactic tides of
the Milky Way Galaxy model. In Section \ref{model}, the Galactic
model, the cluster´s model, the initial conditions, and the
methodology employed are presented. Section \ref{results} shows our
results on the survivability of high-altitude open clusters in the
Galaxy. A brief discussion is presented in Section \ref{discussion}.
Finally, in Section \ref{conclusions}, we present our conclusions.

\section{Galactic tides}
\label{tides}

Spatially extended objects such as open clusters, globular clusters or
satellite galaxies, are affected by the differential gravitational
attraction of their host galaxy (tides). Tidal interactions are
described as the spatial derivative of the gravitational force that an
object exerts on another. Taking for example an open cluster in a
Galactic potential, the acceleration of a star of the cluster at the
position $\textbf{r}$ is given by

\begin{equation}
\label{eq:1}
\frac{d^2\mathbf{r}}{dt^2} = -\nabla \phi_c(\mathbf{r}) -\nabla \phi_G(\mathbf{r}),
\end{equation}
where $\phi_c$ and $\phi_G$ are the gravitational potentials of the
cluster and the host galaxy, respectively.

In order to take the center of mass of the cluster as a frame of
reference, one needs to subtract the acceleration of the cluster's
center of mass due to the host galaxy. The acceleration of a cluster
star at the position \textbf{${\bf r}^{\prime}$} in this frame of
reference is given by

\begin{equation}
\label{eq:2}
\frac{d^2\mathbf{r^{\prime}}}{dt^2} = -\nabla
\phi_c(\mathbf{r^{\prime}}) -\nabla \phi_G(\mathbf{r^{\prime}}) +
\nabla \phi_G(\mathbf{0}).
\end{equation}
In this form equation \ref{eq:2} allows to regroup the last two terms,
and by considering the case where the distance of the star to the
cluster's center of mass is much smaller than the distance of the
cluster to the Galactic center $r^{\prime}\ll R_G$, it is possible to
obtain the linearized equations of motion for a star in the cluster
\citep{2011MNRAS.418..759R} as

\begin{equation}
\label{eq:3}
\frac{d^2\mathbf{r^{\prime}}}{dt^2} = -\nabla \phi_c(\mathbf{r^{\prime}}) + \mathbf{T_t}(\mathbf{r^{\prime}})\cdot\mathbf{r^{\prime}},
\end{equation}
where $\mathbf{T_t}$ is the tidal tensor that contains the effect of
the external potential on the cluster, and its components are given by

\begin{equation}
\label{eq:4}
T_t^{ij}(\mathbf{r^{\prime}}) = \left(-\frac{\partial^2\phi_G}{\partial x^{\prime i} \partial x^{\prime j}}\right)_{\mathbf{r^{\prime}}}.
\end{equation}
Note that the tidal tensor is symmetric ($T_t^{ij} = T_t^{ji}$) and
real-valued, so it can be expressed in its diagonal form. Its
eigenvalues $\lambda$ measure the strength of the tidal field along
the direction of the associated eigenvector. But aside from the
strength, the tidal tensor also contains information about the nature
of the tides produced by the host galaxy. The tide along a given
eigenvector may be compressive ($\lambda < 0$) or extensive ($\lambda
> 0$).  Moreover, the trace of the tidal tensor is directly related to
the local density through Poisson's equation

\begin{equation}
\label{eq:5}
Tr(\mathbf{T}) = \sum_i \lambda_i = -\partial^i\partial^i\phi_G =
-\nabla^2\phi_G = -4\pi G\rho,
\end{equation}
where $\rho$ is the local density, which restricts the trace of
$\mathbf{T}$ to be $Tr(\mathbf{T}) \leq 0$, meaning that at least one
eigenvalue is negative.

This restricts the number of scenarios for the nature of the tide. The
best known example is the case when the total sum is zero, i.e., there
is no external mass inside the volume of interest. In this case, one
or two of the eigenvalues will be positive and the tidal field would
be extensive at least along one direction, this is called extensive
mode. In the scenario we are studying, there will always be Galactic
mass inside the cluster's boundary and $Tr(\mathbf{T})$ will be
negative; thus, in addition to the extensive scenarios, there is the
possibility for all three eigenvalues to be negative at the same time,
meaning that the cluster is compressed along the three directions,
this is called a fully compressive mode \citep{2009ApJ...706...67R}.

\section{Galactic model and Numerical Implementation}
\label{model}
The main goal of this work consists of a statistical study of open
clusters destruction, specifically by tides, within a detailed and
realistic model of the Milky Way's disk. Such mechanism is effective
for clusters that experiment lifts away from the disk plane. Each
passing through the disk would be a violent event that may compromise
their survival. In our simulations we place several of these objects
at different galactocentric radii, different heights above the plane
and different initial masses of the clusters. For the purpose of this
study, a very detailed Milky Way potential is required, along with a
good numerical implementation that is not computationally burdensome
for each cluster.

Although $N$-body simulations seem to be the perfect tool to study
this type of problems, they are not suitable to achieve the goals of
this work yet. First of all, the potential model employed is a very
detailed steady model adjusted specifically to the best of recent
knowledge of the Galaxy structural and dynamical parameters (i.e.,
spiral arms and bar angular velocities, masses, density laws,
scale-lengths, etc.). Unlike $N$-body simulations, the potential we
use is totally adjustable; we are able to fit the whole axisymmetric
and non-axisymmetric potential (i.e., spiral arms and bar), in three
dimensions to our best understanding of any particular galaxy from
observations and models.

Secondly, our model is considerably faster computationally speaking
than $N$-body simulations. This allows us statistical studies by
swapping the structural and dynamical parameters. At the same time we
can study in great detail individual stellar orbital behavior such as:
vertical structure, resonant regions, chaotic {\it vs.} ordered
behavior and periodic orbits to estimate at some degree orbital
self-consistency, etc., without having the customary resolution
problems of $N$-body simulations. We achieve this without resorting to
simple {\it ad hoc} models for a spiral perturbation (like the widely
employed cosine potential) or the bar (like a Ferrers bar); instead,
we employ a three-dimensional (3D) mass distribution for the spiral
arms and a bar that fits the Milky Way's density bar, from which we
derive their gravitational potential and force fields \citep[for a
  thorough description of the model see][]{PMME03,Pichardo2004}. We
introduce briefly here the Galactic model and a detailed description
of the numerical implementation to describe open clusters and their
interaction with the Galaxy.

\subsection{Milky Way's Gravitational Potential}
The model includes an axisymmetric potential, formed by a
\citet{1975PASJ...27..533M} disk and bulge, and a massive halo
\citep{Allen1991}.  For the spiral arms we employ the PERLAS model
\citep{PMME03} that consists of a bisymmetric three-dimensional
density distribution. For the Galactic bar we use a non-homogeneous
triaxial ellipsoid that reproduces the density law of the COBE/DIRBE
triaxial central structure of the Galaxy. For further details on the
model see \citet{PMME03,Pichardo2004}.

To fit our model we make use of observational / theoretical parameters
from the literature. The length of the bar (semi-major axis) is set to
3.5 kpc, with scale-lengths of 1.7, 0.64, and 0.44 kpc. The total mass
is 1.4 $\times$ 10$^{10}$ M$_{\odot}$ with a pattern speed $\Omega_B$
= 45 $\kmskpc$. For the spiral arms, we consider a pitch angle of $i$
= 15.5$^{\circ}$ and a mass ratio of $M_{\rm arms}/M_{\rm disc}$ =
0.05 at a pattern speed $\Omega_S$ = 20 $\kmskpc$. Further details on
the parameters of the model and observational restrictions considered
were introduced in Table 1 of \citet{2016MNRAS.463..459M}.

\subsection{Tidal tensor approach}
\label{TidalTensor}

Regarding the interaction of the cluster with the Galaxy, this can be
implemented mainly in two ways: in a direct approach the cluster is
placed within the Galactic model, and the acceleration of a star is
the sum of the acceleration due to the rest of the cluster and the
acceleration due to the Galaxy. Although this approach seems natural,
the size of the two systems, cluster and galaxy, differs by orders of
magnitude, so that there would be two very different time steps in the
simulation. This would increase severely the computation time and may
even lead to numerical errors \citep{2015MNRAS.448.3416R}. In the
tidal tensor approach the equations of motion of a star are solved in
the frame of reference of the cluster's center of mass (equation
\ref{eq:2}). This means that, as seen in Section \ref{tides}, the
contribution of the Galaxy to the equations of motion is contained in
the tidal tensor (equation \ref{eq:3}). This method, that captures
very well the interaction of the cluster with the Galaxy, has been
tested in several works by comparing with the direct approach,
ensuring its validity and accuracy
\citep{2011MNRAS.418..759R,2013MNRAS.436.3695R}.

Since close encounters with individual stars are unlikely to be
important and all large structures of the Galaxy have sizes much
bigger than the size of a cluster, a linearized tidal tensor approach
is ideal (i.e., sufficient and relatively fast) for this study. For
our purposes, we need to compute the tides at every point of the
pre-calculated cluster's orbit. The computation of the tidal tensor
along the orbit is as described in Section \ref{tides}, and the total
acceleration of every star is the sum of the acceleration due to the
rest of the cluster and the tidal acceleration (equation \ref{eq:3}).

\subsection{Cluster modeling}
\label{cluster}

Since we attempt to cover all the possible scenarios that result from
different combinations of initial galactocentric radius, orbital
height, and cluster's mass, our study requires hundreds of
simulations; this compounded with the complexity of the Galactic
potential, makes it an impossible task for an $N$-body code. This is
because in the $N$-body approach particles interact with one another
directly, and an accurate representation of the gravitational field
will depend on the number of particles, N, which for practical
purposes depends on the computational capabilities available. Also the
two-body relaxation is a well-known effect unavoidable in $N$-body
simulations.

For these reasons, and for the purposes of this work, it is desirable
to use a more efficient technique that allows us to approach the
gravitational potential of the system without computational
restrictions, and isolate the effects of the interaction between the
cluster and the Galaxy, avoiding two-body relaxation.

In this work we use the Self-Consistent Field method (SCF)
\citep{1972Ap&SS..16..101C,1973Ap&SS..23...55C,1992ApJ...386..375H} to
simulate each N point-mass particle cluster. The method approximates
the true gravitational potential and density of the cluster with a
finite series of basic functions, i.e., particles in SCF techniques do
not interact with one another directly but only through their
contribution to the gravitational field of the system.

SCF techniques are truly efficient when the number of basic functions
is kept small, which is only possible when the shape of the system
does not deviate much from a given symmetry. In the most perturbed
scenarios, the gravitationally bound part of stellar clusters may
display spheroidal shapes, but for most of the cases they do not
deviate significantly from sphericity \citep{1999A&A...352..149C},
mainly because the parts of the cluster more prone to perturbations
are the less dense outer regions, instead of the more massive inner
region \citep{2006MNRAS.367..646L}.

 Assuming that the distribution of particles is concentrated, the
 gravitational potential in the outer parts will be nearly spherical
 \citep[which is the case for more than half of the open
   clusters,][]{2002A&A...383..153N} and following the SCF technique
 for the computation of the gravitational potential, we use a basis
 whose first member is the Plummer density-potential pair
 \citep{1973Ap&SS..23...55C}; in other words, at each time step in the
 simulation we account for self-gravity within the cluster, by
 assuming that the gravitationally bound particles are embedded in a
 Plummer potential. The criteria to decide the membership of a given
 star to the cluster is that the star's velocity, $v$, must be less
 than the local escape velocity, $v_e$. Those stars with $v \geq v_e$
 do not contribute with mass to the cluster at that time step of the
 simulation (note that the velocity of such star could diminish at a
 later time, $v < v_e$, and it would once again contribute to the mass
 of the cluster), in this manner the mass of the system evolves with
 time. Also we let the radial scale in the background Plummer
 potential to evolve with time by computing, at each time step, the
 half-mass radius, $r_h$, of the distribution, that for the Plummer
 profile is directly related to the radial scale-length, $a$, by $r_h
 \approx 1.3a$. In this way the shape of the cluster, although assumed
 to be always a Plummer sphere, is also evolving with time, with the
 ability of mimicking expansions and contractions of the system.

For the initial conditions we distribute the particles in space
according to a Plummer density profile. The velocity space is
populated by using the rejection technique with a Plummer distribution
function, in this way the initial distribution of test particles is
relaxed with the background potential.

We simulate clusters within galactocentric radii, 5.5 kpc $\la$
  $R_{min}$ $\la$ 14 kpc, with vertical departures from the disk
  plane, 0 kpc $\la z_{max} \la$ 5.5 kpc and masses, $500 {\rm
    M}_\odot \la M_0 \la 5 \times 10^4 {\rm M}_\odot$.

With this numerical approach we do not attempt to reproduce all the
physics of the internal mechanisms involved in the evolution of a star
cluster; instead, by avoiding the mass loss due to processes intrinsic
to the cluster, we focus solely on its interaction with the host
galaxy, that in the case of clusters suffering disk shocking is of
greater importance in magnitude.

\section{Results}
\label{results}

\subsection{Milky Way's tidal field}

After building the model for the Milky Way's gravitational potential,
a computation of the strength of the tides in every point of space
will provide us a general insight about the tidal interaction of a
given open cluster with the Galaxy, depending on its position, i.e.,
its orbit through the Galaxy. The tidal field of the model will also
give us information about the nature of the tidal modes (extensive or
compressive), and of particular interest is the tidal field of the
spiral pattern and the central bar.

To compute an x-y projection of the tidal field of the model we locate
a mesh of $256\times256$ points on the plane. The first thing to do in
order to obtain the strength of the tide at every point on the mesh,
is to compute the nine components of the tidal tensor at every point
(equation \ref{eq:4}). Since the tidal tensor is given as the second
spatial derivative of the potential, which is not totally analytical
for our model, we compute its components with a first-order difference
scheme.  This is done by placing six additional points (two along each
direction) around the point of interest, and computing the
acceleration at each one of these points. By subtracting the
accelerations of every pair of points along the three directions
(according to the difference scheme) we obtain the nine components of
the tidal tensor on the entire Galactic plane; note that the tensor is
expected to be symmetrical but, since the force is analytical and the
derivative is numerical, the cross terms are not necessarily
identical, in fact they differ in one part in a million, proving the
quality of our derivatives. Next, the tensor is diagonalized to obtain
the three eigenvalues; its amplitude and sign give us the strength and
nature of the tides, respectively.

\begin{figure}
\begin{center}
\includegraphics[width=9cm]{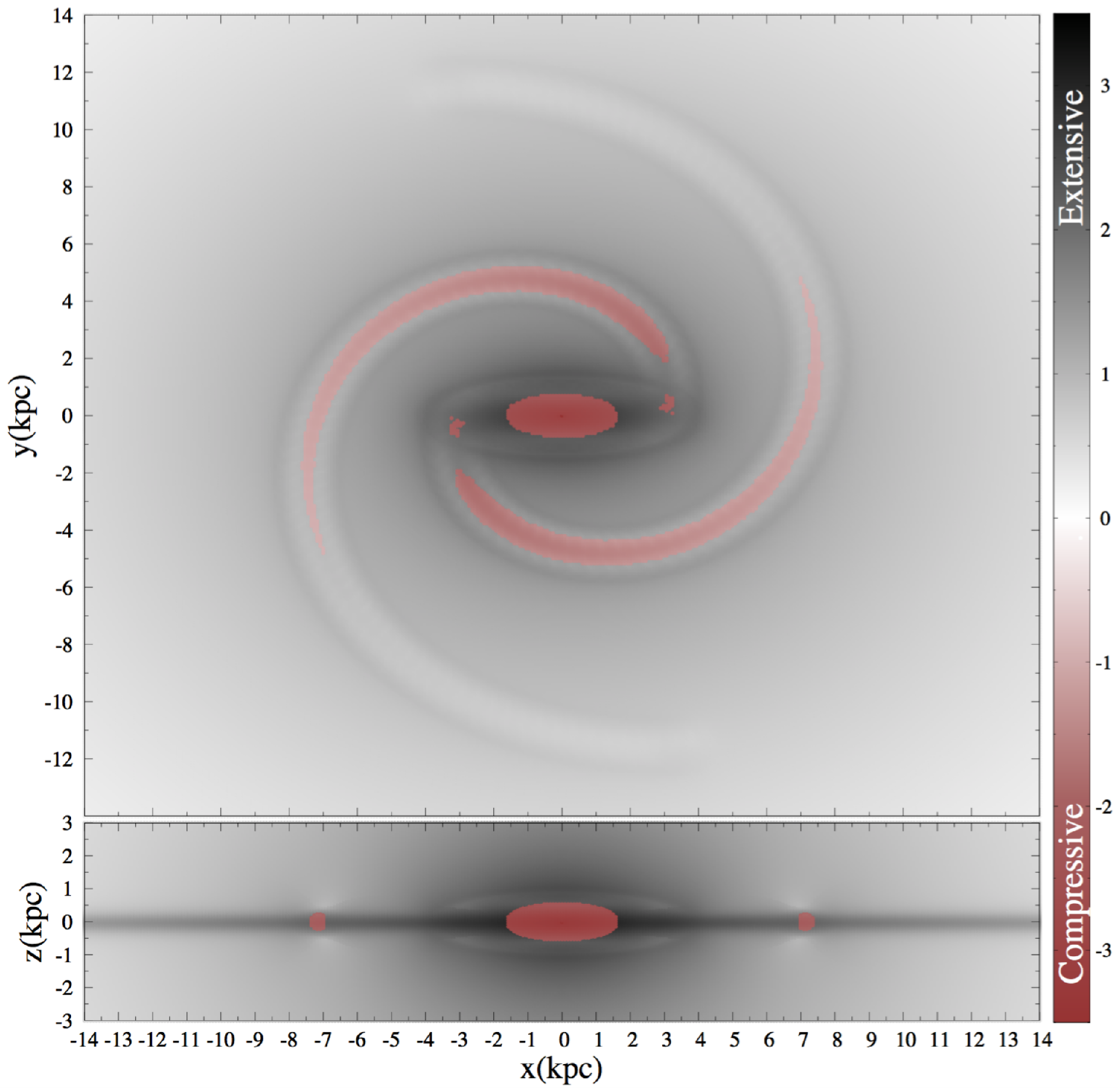}
\end{center}
\caption{Tidal Field of the Galaxy model. The color map indicates the
  strength of the tides and the regions with extensive or compressive
  tidal modes. Compressive regions are located at the center of the
  bar and the spiral arms.}
\label{fig:Tidalmap}
\end{figure}

The upper part of Figure \ref{fig:Tidalmap} shows the $x-y$ projection
of the tidal field measured as $\sqrt{\lambda_1^2+\lambda_2^2}$, where
$\lambda_1$ and $\lambda_2$, are the eigenvalues corresponding to the
eigenvectors that lay on the Galactic plane. The color map shows that,
on the Galactic plane, the tides are stronger near the Galactic center
and become weaker at larger radii. By marking the points where all the
eigenvalues of the tidal tensor are negative, is interesting to notice
that the center of the bar and the center of the spiral pattern, all
along the arms, are dominated by fully compressive tides. Also notice
that for the spiral arms, when we move along the transversal
direction, the tides change from being fully compressive to dropping
to zero, then becoming extensive at the border of the arms.

The existence of regions with fully compressive tides within the bar
and spiral arms, gives to these structures an important role on the
survival of open clusters that move on the Galactic plane. By its
nature, the fully compressive tides will enhance the binding of the
stellar cluster, preventing it from being disrupted as would happen if
the cluster moves to other regions of the Galaxy where the tides are
extensive; this means that clusters that are born in the spiral arms
or the bar will be tidally protected from destruction all the time
they stay there. In a preliminary study, we placed several 500 solar
mass clusters at a galactocentric distance of 7 kpc at different
angular positions; we found that the survival time for those clusters
born outside the compressive region of the spiral arms ranged between
34 and 50 Myr, while a cluster that was born and lived inside the
compressive region of the spiral arm, survived 70 Myr, i.e. a cluster
inside the compressive region of the arm lived almost twice as long as
the ones outside. It should also be noted that the orbit of this
cluster only laid inside the compressive region of the spiral arm for
80 Myr, but this time depends strongly on the chosen galactocentric
radius. In a future work we will present a detailed study of the
implications that the fully compressive tidal modes of the bar and
spiral arms have on the survival of open clusters in galaxies.

The lower part of Figure \ref{fig:Tidalmap}, is similar to the upper
part but it is the $x-z$ projection of
$\sqrt{\lambda_1^2+\lambda_3^2}$. It shows a transversal cut of the
bar and spiral arms, we also see regions of fully compressive
tides. But the tidal field is dominated by the extensive modes due to
the Galactic disk. In the next sections we will study the interaction
of open clusters with the tidal field of the Galaxy, particularly for
high-altitude open clusters whose orbits take them to cross the
Galactic disk several times.

\subsection{Open Clusters in the Galactic Plane}
The numerical procedure described in Section \ref{cluster} allows
us to compute the disruption rates of an open cluster in its orbit
through the Galaxy: as it moves on the Galactic plane, interacts with
the spiral arms, and when its orbit takes it to make vertical
excursions above and below the Galactic disk.

The fact that the disruption time depends on the orbital
galactocentric distance is a very well known result
\citep{2011MNRAS.418..759R,2012MNRAS.419.3244B}. In order to validate
our numerical implementation to simulate clusters and isolate their
tidal interaction with the Galaxy, we start with open clusters moving
in the Galactic plane.

All clusters have an initial half-mass radius $r_h=10$ pc. We explore
different mean galactocentric radii for the orbit and different
initial masses of the cluster.

\begin{figure*}
\begin{center}
\includegraphics[width=18cm]{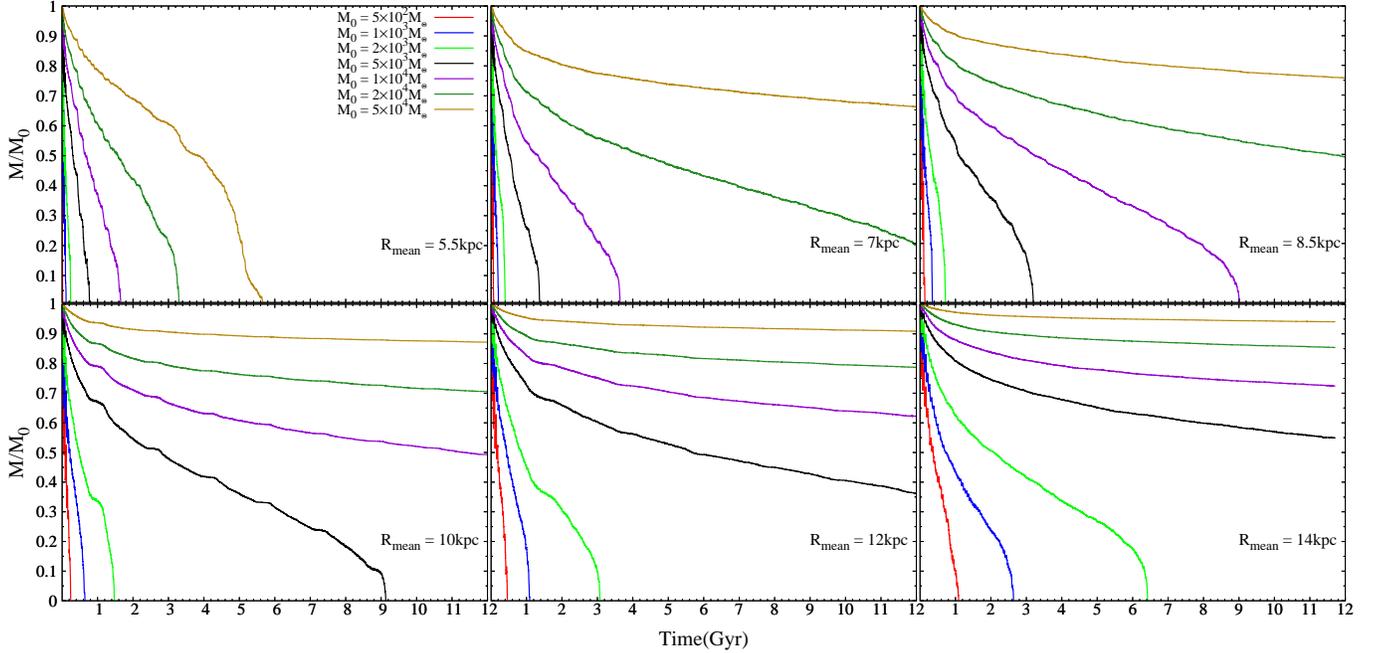}
\end{center}
\caption{Evolution of the cluster's mass, normalized to its initial
  value, for different galactocentric radii of the orbit. Each color
  indicates different initial masses of the cluster, ranging from
  500 M$_{\odot}$ to $5\times10^4$ M$_{\odot}$.}
\label{fig:OnPlane}
\end{figure*}

Figure \ref{fig:OnPlane} shows the dependence of the survival time of
the cluster on the environment (galactocentric radii) and the initial
mass. As anticipated from the tidal map of the Galaxy (Figure
\ref{fig:Tidalmap}), the lifetime of all simulated clusters increases
as they move away from the Galactic center, this is because the
strength of the tides decreases with distance. Regarding the
dependence on the initial mass, at all galactocentric radii, clearly
the initially more massive clusters will survive longer; some of them
even for the age of the Universe if they move in orbits far away from the
Galactic center. Also, as a consequence of the small tidal strength at
large radii, even non-massive clusters could survive for more than 1
Gyr as long as they move in the outer regions of the disk.

Regarding the clusters that can be totally disrupted, within the
lifetime of the Galaxy, by the tidal field produced by the
non-axisymmetric structures in the Galaxy, we can try to determine
their lifetime. For clusters with 2,000 ${\rm M}_\odot \la M_0 \la
\ 10,000 \ {\rm M}_\odot$ and 5.5 kpc $ \la R_{\rm mean}\la$ 8.5 kpc
(all orbits with low eccentricities to assure they lay very close to a
chosen galactocentric radius), the survival time can be parameterized
as:
\begin{eqnarray}
\nonumber
\log\tau_0(R_{\rm mean}) &=& -3.1042-0.2628 R_{\rm mean} + 0.4950 \log(M_0) \\ 
&& +0.1268 R_{\rm mean} \log(M_0)
\end{eqnarray}
where $\tau_0$ is expressed in Gyr, $M_0$ in M$_\odot$, and $R_{\rm
  mean}$, the mean galactocentric radius is expressed in kpc.  This
fit has a typical deviation of $1.5\%$ with our simulations and it is
always better than $3.5\%$.

Intrinsic processes of open clusters may impose further restrictions
and better limits to the destruction rates. Our study however, shows
the correct behavior of the curves representing the mass loss of the
cluster (Figure \ref{fig:OnPlane}), which adopt a similar shape than
those from works using $N$-body models
\citep{2000ApJ...535..759T,2003MNRAS.340..227B,2013ApJ...778..118W};
i.e., the results of our numerical approach to tidal disruption
resemble those of $N$-body codes and give us a good estimation for the
survival time of clusters immersed in the tidal field of the Galaxy.

\subsection{High-altitude Open Clusters: Crossing the Galactic Plane}
While preparing this work, an interesting paper recently accepted,
employing $N$-body simulations \citep{2016arXiv160802309R} found that
for open clusters experiencing large excursions from the Galactic
plane, the effect of tides due to crossings with the disk is
negligible. However, it should be noticed that, in the majority of
cases, $N$-body simulations have severe problems to resolve the thin
massive disk of galaxies, this means that tidal shocks are likely to
be critically underestimated.

Several studies in the literature deal with the evolution of open
clusters within a Galactic model. Depending on the mechanisms being
studied, which usually are internal to the cluster, the external
potential can be simplified as much as needed
\citep{2014MNRAS.437..916G,2014MNRAS.444...80O,2016MNRAS.455..596C};
keeping the external potential simple allows focusing on an specific
internal mechanism. Another reason to simplify the galactic potential
is the fact that the implementation of a realistic galactic model into
an $N$-body code becomes too complicated as well as computationally
expensive.

With our adopted numerical implementation to account for the
self-gravity of the cluster, and given that the tidal tensor approach
captures the contribution of the Galaxy's gravitational field along
any orbit, we can place the simulated clusters on more general orbits,
no longer restricted to the plane.  This method allows us to compare
the survival times of clusters moving in the plane of the Galactic
disk with those that move above and below, crossing the disk. This
will establish the role that the tides due to the disk, which are
dominant in the $z$ direction (Figure \ref{fig:Tidalmap}), have on the
cluster's disruption.

\begin{figure}
\begin{center}
\includegraphics[width=9cm]{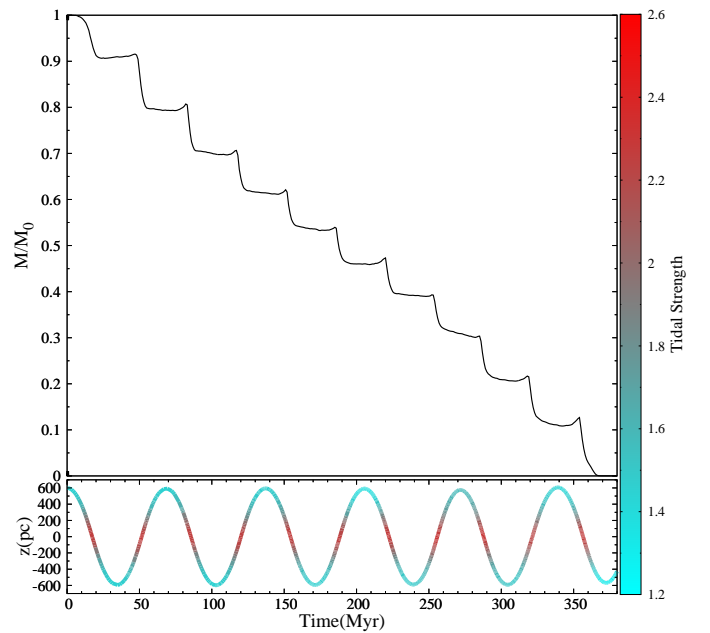}
\end{center}
\caption{Top: Evolution of the cluster's mass, normalized to its
  initial value, for a cluster that crosses the plane of the disk
  several times along its orbit. Bottom: Evolution of the cluster's
  vertical separation from the plane. The color coding indicates the
  strength of the tides at that point of the orbit.}
\label{fig:TidalOrbit}
\end{figure}

Upper panel of Figure \ref{fig:TidalOrbit} shows the mass evolution
for a cluster initially placed at a galactocentric radii $R=5.5$ kpc
and height $z=600$ pc above the Galactic plane. Moving along that
orbit, the cluster experiences sudden losses of mass, and its
evolution is far from being smooth. Bottom panel of Figure
\ref{fig:TidalOrbit} shows the evolution of the cluster's vertical
separation from the plane. The color coding, that indicates the
strength of the tides at that point of the orbit, shows what we saw in
Figure \ref{fig:Tidalmap}: that the strongest tidal interaction will
occur when crossing the disk. By comparing the two panels in Figure
\ref{fig:TidalOrbit} we can see that every sudden loss of mass occurs
just when the cluster crosses the Galactic plane. This result shows
how important the vertical displacements are, and that the cluster can
easily be disrupted, even when it spends less time interacting with
the material of the disk.

We explored several other values of the maximum vertical distance
reached by the cluster in order to compare with those that move only
in the plane, as well as to quantify how the cluster's survival time
depends on its maximum orbital height.

Next we present a series of simulations for clusters placed at
different orbits. These orbits differ in the maximum height reached
but, for all of them, its projection on the plane has the same
$R_{mean}$.

\begin{figure}
\begin{center}
\includegraphics[width=9cm]{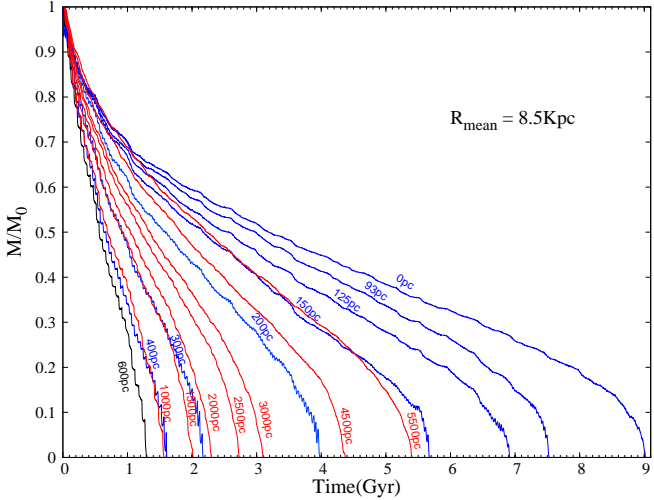}
\end{center}
\caption{Mass evolution, normalized to its initial value, for clusters
  moving in different orbits. The maximum orbital height is indicated
  on each curve and goes from 0 (motion confined to the Galactic
  plane) to 5500 pc. Blue lines indicate the range of orbital altitudes
  for which the lifetime of the cluster decreases as the altitude
  increases. Red lines indicate the regime for which the lifetime of
  the cluster increases as the altitude of the orbit increases.}
\label{fig:MassvsTime}
\end{figure}

Figure \ref{fig:MassvsTime} shows the mass evolution for clusters that
reach different orbital heights, ranging from 0 to 5500 pc, all of
them at $R_{mean}=8.5$ kpc. The evolution of the mass is not equally
smooth for all clusters as it depends on the orbital height. Also the
lifetime of the cluster varies depending on the orbit.

Figure \ref{fig:MassvsTime} shows that a cluster survives more time if
it is confined to the Galactic plane, without vertical
excursions. Then, the lifetime decreases as the orbital height
increases. But this trend does not continue forever, it is reversed
when the orbital height reaches a turnaround point of about 600 pc
($\sim2.5$ times the vertical scale of the disk), from there the
lifetime of the cluster increases as the orbital height increases.
Here we found a correlation between lifetime and orbital height, that
exhibits two different behaviors separated by a turnaround height. In
the next section we will analyze this result in detail.

\subsection{The Lifetime of High-Altitude Open Clusters}
\begin{figure}
\begin{center}
\includegraphics[width=9cm]{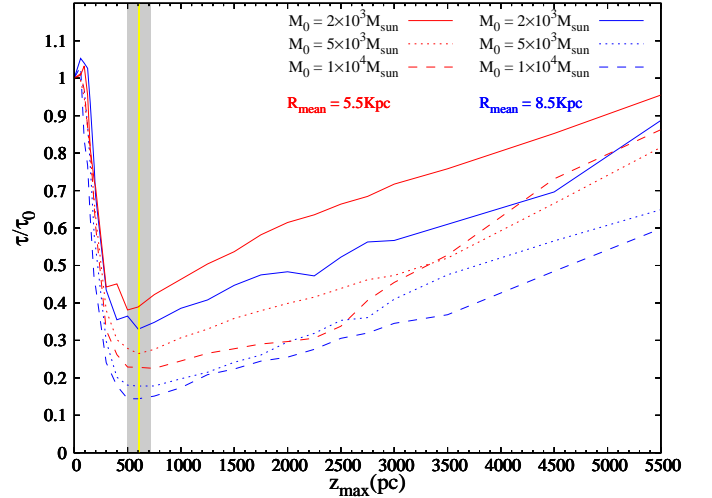}
\end{center}
\caption{Lifetime of each cluster, $\tau$ (normalized to the lifetime
  of the cluster that moves on the plane, $\tau_0$) as a function of
  the maximum orbital height $z_{max}$. Red and blue lines correspond
  to orbits with galactocentric distances of $5.5$ kpc and $8.5$ kpc,
  respectively. Three different initial masses are presented for each
  galactocentric radius. The curves reach their minimum survival time
  at slightly different values of $z_{max}$, the mean of this values
  occurs at $z_{max} \approx 600$ pc (yellow line). The shaded band
  encloses the minima for all the curves.}
\label{fig:TimevsHeight}
\end{figure}

In Figure \ref{fig:TimevsHeight}, we analyze in detail the behavior of
the correlation between lifetime and orbital height, derived in the
previous section; it is a plot of the maximum orbital height,
$z_{max}$, {\it vs.} the lifetime of each cluster, $\tau$ (normalized
to the lifetime $\tau_0$ of a cluster with the same mass and
$R_{mean}$ that moves on the plane of the Galaxy). We have included a
wide exploration of values for the maximum height reached by the
orbit, ranging from 0 to 5500 pc, for clusters at two different
galactocentric distances and with three different initial masses.

Again, we see that the clusters that survive longer are the ones that
move in the plane or remain very close to it, up to 125 pc ($\sim0.5$
times the vertical scale of the disk). As the height of the orbit
increases, the lifetime decreases until the curve reaches its minimum
at some value of $z$, we will call this the turnaround point.

This behavior reveals two distinct regimes in the evolution of
clusters whose orbits take them to cross the disk.  In the first
regime the clusters that live longer are the ones that remain within
the denser part of the disk, i.e., those in the plane or with small
vertical excursions. Then, for orbits that reach higher altitudes the
cluster will cross the disk at a high velocity, which means that the
vertical gravitational force acting on the cluster varies rapidly with
time, and as a consequence also the tidal strength acting on the
cluster.  This leads to a considerable loss of bound stars, and hence
an important loss of mass each time the cluster crosses the plane of
the disk.

In the first regime the lifetime of the cluster starts to decrease
rapidly as the crossings through the plane become more violent, then
slows down until it reaches a minimum for a given value of
$z_{max}$. From this point, and for higher altitudes, the lifetime of
the cluster increases as it raises above the Galactic plane; in this
second regime, although the cluster falls towards the disk and crosses
it at high velocity, the frequency of its vertical motion diminishes
as the orbital height increases, which means that the number of
crosses through the Galactic plane is less compared with clusters
placed at lower altitudes. Therefore, the cluster interacts less with
the tidal field of the disk increasing its lifetime.

The survival time as a function of height away from the plane also
depends on the average galactocentric radius and the mass of the
cluster. This can be parameterized as:
\begin{equation}
\label{taufit}
{\tau(z)\over\tau_0}= \left[\alpha + \beta z + \gamma z^2 \right]+\left[{(1- \alpha) \over 1+\exp\left({z^{1/8}-\delta \over 0.02544}\right)}\right];
\end{equation}
where the parameters $\alpha$, $\beta$, $\gamma$, and $\delta$, have a
small dependence on the galactocentric radius and the initial mass of
the cluster; the functional form of each parameter can be expressed as
\begin{equation}
\chi(R_{\rm mean},M_0)=\chi_0+\chi_R(R_{\rm mean}-7)+\chi_M[\log(M_0)-3.6666],
\end{equation}
where $R_{\rm mean}$ is in kpc, $M_0$ in M$_\odot$, and the values of
the $\chi_0$, $\chi_R$, and $\chi_M$ are presented in Table
\ref{tabla1}. The rms difference between this fit and our simulations
is $6.3\%$.

\begin{deluxetable}{lccc}
\tablecaption{Parameters used to estimate survival time as a function of height
\label{tabla1}   }
\tablewidth{0pt}
\tablehead{
\colhead{$\chi$} & \colhead{$\chi_0$} & \colhead{$\chi_R$} & \colhead{$\chi_M$}
}
\startdata
$\alpha$ & 0.21371 & -0.02746 & -0.24042\\
$\beta$ & 0.07250 & 0.00190 & -0.00539 \\
$\gamma$ & 0.00610 & -0.00168 & 0.00083\\
$\delta$ & 0.81451 & -0.00146 & -0.0079
\enddata
\end{deluxetable}

Notice that the first term in equation \ref{taufit}, represents the
second regime (i.e., high velocity, low frequency plane crossing
regime); while the first regime is represented by the second part of
the equation plus the constant $\alpha$.

In summary, the disruption of open clusters due to the tidal field of
the disk will depend on the velocity at which the cluster crosses the
Galactic plane and the number of times this occurs. In a first regime,
for altitudes between 0 and 600 pc ($\sim2.5$ times the vertical scale
of the disk) the crossing velocity will be the main disruption factor,
with slower crosses being gentler. In a second regime, for altitudes
higher than 600 pc it will be the number of crosses through the disk,
not the velocity, what will determine the lifetime of the
cluster. While in the pioneer work by \citet{1972ApJ...176L..51O},
applied to globular clusters (very high altitude case), the
destruction of clusters is proportional to the velocity, $v_z^{-2}$,
while in our simulations, for the second regime, we find that the
destruction is nearly independent of the velocity ($\propto
v_z^{-0.2}$) and only proportional to the number of crossings; this is
probably related to the fact that clusters in our simulations are
moving too slow and are too small for the impulse approximation to be
valid.

In any case, for the masses and orbital heights explored in our
simulations, the clusters that survive the longest are those that move
very close to the plane of the disk.

\subsection{Survival of lifted clusters}
In the previous sections we studied two different survival scenarios
for open clusters in the Galaxy: clusters that move all their life in
orbits very close to the Galactic plane, or clusters with orbits that
oscillate away from the Galactic plane. But, what happens in the
transition of clusters from one environment to the other? i.e., what
will be the evolutionary path followed by a cluster that is born and
lives the first stages of its life in the Galactic plane and
afterwards it is lifted by some mechanism above the plane?

In a previous work \citep{2016ApJ...817L...3M}, we showed that a
mechanism able to lift open clusters way up the Galactic plane are the
spiral arms. The presence of the spiral arms induces several clusters
to be violently lifted from the Galactic plane, to heights up to 3
kpc.  An important question for such mechanism is whether the cluster
would survive such lifting. For this test, we have selected some
high-altitude orbits, i.e., clusters that lived first for a while in
the plane and then rose abruptly.

\begin{figure*}
\begin{center}
\includegraphics[width=18cm]{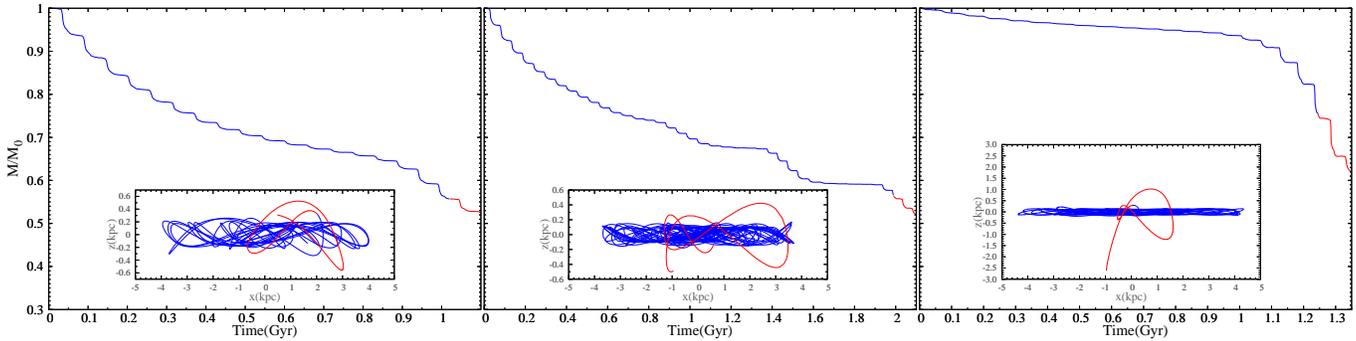}
\end{center}
\caption{Evolution of the cluster's mass, normalized to its initial
  value, for clusters lifted above the Galactic plane.  The insets
  show the $x$-$z$ projection for the orbit of each cluster. The blue
  path indicates the part of the orbit confined to the plane, while
  the red path indicates the motion and transition of the cluster into
  higher altitudes.}
\label{fig:Orbits}
\end{figure*}

In Figure \ref{fig:Orbits} we show the mass evolution, as well as the
orbits, for three examples of lifted clusters from the work by
\citet{2016ApJ...817L...3M}. The blue lines indicate the life stage of
the cluster when it moves confined to the Galactic plane; the red
lines indicate the second part of its evolution: when the cluster
oscillations take it away from the plane, crossing the disk.

Note that the steps in the blue line, in the mass evolution track, are
due to the eccentricity of the blue path, while the steps in the red
line are not only due to the eccentricity but also due to the
crossings of the disk. Nonetheless, in spite of these two disruption
factors (orbital eccentricity and disk shocking), we find that several
clusters can make the transition from moving in the plane to be
located at high altitudes surviving the sudden acceleration
associated with the lifting.

\section{Discussion} \label{discussion}
It is interesting to note that clusters in the Galactic plane, while
moving along the transversal direction of an arm, feel the effect of
very different environments in a short period of time: they go from
moderately extensive tides (outside the arm), to very extensive, to
lightly extensive, to mildly compressive tides (middle of the arm), to
lightly extensive, very extensive, and finally to moderately extensive
(outside the arm again) in the crossing of an arm. 

The compressive nature of the inner region of the arms and the bar has
an important role on the survival of small clusters in the disk by
enhancing the binding of the clusters, protecting them from
disruption: clusters that live in the spiral arm region survive longer
within the spiral arms than in any other place of the Galaxy.
Additionally, note that colder material of the Galaxy, like new born
clusters, would remain longer in the spiral arm region, i.e., an
overdensity in the center of the spiral arms of a factor of 4/3
\citep[built in our model from observations,][]{2011MNRAS.418.1423A},
results in a slowing down of the clusters, in the spiral arm region,
by a factor of 3/4; this means that an overdensity of 36$\%$ (could be
even larger in the Milky Way case) represents a slowing down of
1-(1/1.36) = 27$\%$ by the clusters on their trajectories with respect
to the spiral pattern. For us it also means an enhancement of 36$\%$
on the birth rate of the clusters. This implies that a cluster born at
7 kpc inside the spiral arms region, considering the Galactic
rotation, shape of the orbits, and arms density contrast, would spend
up to 80 million years within the compressive zone of the arm, which
means that clusters as massive as 500 M$_\odot$ would be more abundant
in the spiral arm zone, not only because of the density contrast or
because of an enhancement of the star formation, but because of the
binding by tides. In a preliminary study, we have calculated the
survival time for clusters born outside the compressive region of the
spiral arms and found that those are ranged between 34 and 50 Myr,
while the cluster that was born and lived inside the compressive
region of the spiral arm, survived 70 Myr, i.e., the cluster inside
the compressive region of the arm lived almost twice as long as the
ones outside.

A few mechanisms are known to explain the existence of open clusters
at high altitudes away from the disk plane, but once the clusters are
lifted the possibilities of survival are different from the ones in
the plane. On one hand, it is assumed that clusters living distant
from the Galactic plane have better prospects of surviving as a unit
since they live far away from the destructive effects of in-plane
substructure, such as molecular clouds, the spiral arms, and the bar.
There is, however, a well known destructive mechanism that is, in
general, ignored in reference to open clusters: tidal disk shocking;
studies in this field ignore it since the majority of open clusters
spend all their life as a bound entity in the disk plane, however this
is not the case for approximately 13\% of clusters
\citep{2008ApJ...685L.125D}. On the other hand, some work has found
this mechanism to be negligible when studied in $N$-body simulations
\citep{1998MNRAS.299.1019V,2016arXiv160802309R}, highly likely due to
difficulties to vertically resolve the thin massive Galactic disk. Our
approach can in fact resolve the thin Galactic disk (as well as many
other components of the Galaxy).

With this technique we find that for high-altitude open clusters,
crossing the disk plane can have a severe tidal destructive effect, up
to six times as destructive as effects on the disk plane such as the
spiral arms and the bar. Two regimes are displayed based on their
maximum altitude; in the first one, for clusters with $z_{max}\la$600
pc, the lifetime diminishes as the orbital altitude (as well as the
strength of the planar shock) increases, i.e., clusters confined to
the Galactic plane would live longer than those with orbits that take
them a few hundred parsecs away from the plane because they would not
experience vertical tidal shocks from the Galactic disk. In the second
regime, for clusters that orbit higher than $\sim$600 pc, lifetimes
increase again because, although tidal shocks keep being powerful, the
frequency of encounters with the disk declines as the altitude
increases (i.e., for two orbits with different $z_{max}$, the number
of crossings with the disk, in the same period of time, is different).

Finally, for clusters in the transition from the thin disk to the
thick disk, lifted by the effect of the spiral arms, we find that they
survive the sudden acceleration.

\section{Conclusions} 
\label{conclusions}
With the use of a very detailed observationally motivated model of the
Milky Way Galaxy, we produced an ensemble of numerical simulations of
the effect of Galactic tides on open clusters in different regions of
the Galaxy and at different heights away from the disk plane.

We construct a tidal map from our Milky Way model. We find important
zones where the tides are fully compressive within the bar and spiral
arms.  By their nature, fully compressive tides would magnify the
binding of stellar clusters, preventing them from being disrupted as
fast as it would happen for clusters moving into any other region of
the Galaxy. 

Fully compressive regions have interesting implications in the
evolution of clusters within the Galactic plane, specially on the time
small clusters last inside the spiral arms in comparison to other
locations of the Galaxy, i.e., small clusters would be more abundant
in the spiral arm zone, not only because of the density contrast or
because of an enhancement by the star formation, but because of the
protecting binding by tides. Preliminary calculations show that
clusters inside the compressive region of the arm live almost twice as
long as the ones outside.

Contrary to common knowledge, high-altitude open clusters (those over
200 pc) experience severe tidal destruction when they pass through the
disk. In this respect, we find that, up to a certain altitude away
from the disk plane (between 0 and 600 pc), the lifetime of clusters
decreases as its maximum orbital altitude increases because tidal
shocks with the disk are stronger; clusters confined to the Galactic
plane live longer because they do not experience vertical tidal shocks
from the Galactic disk (even in the presence of strong arms and
bar). For clusters that raise higher than $\sim 600$ pc, lifetimes
increase again, as a function of the maximum altitude, due to the
fewer number of encounters with the disk.

We parameterized the survival time for high-altitude open clusters
(equation \ref{taufit}), as a function of height away from the plane,
average galactocentric radius and the mass of the cluster.

Finally, for clusters in transitional orbits (those that evolve from
the thin disk to the thick disk due to the effect of the spiral arms),
we find that they experience a sudden acceleration that although
violent, is not strong enough to totally destroy them.

\acknowledgments We thank the anonymous referee for a careful reading
and suggestions that greatly improved this paper. We acknowledge the
support of DGTIC-UNAM for providing HPC resources on the Cluster
Supercomputer Miztli. We acknowledge DGAPA-PAPIIT through grants
IN-114114, IN-105916, IN-109716 and IG-100115. L.A.M.M. acknowledges
support from DGAPA-UNAM postdoctoral fellowship.

\end{document}